\documentclass[preprint,review,12pt,authoryear]{article}

\usepackage{graphicx}
\usepackage{natbib}
\usepackage{amssymb}
\usepackage{amsmath}

\begin{document}

\title{Reconciling the Observed Mid-Depth Exponential Ocean Stratification with Weak Interior Mixing and Southern Ocean Dynamics via Boundary-Intensified Mixing}

\maketitle

\begin{center}
\author{Madeline D. Miller\footnote{corresponding author, now at MIT Lincoln Laboratory, Lexington, Massachusetts, USA, madeline.miller@ll.mit.edu}, Xiaoting Yang, Eli Tziperman\\~\\
Department of Earth and Planetary Sciences and School of Engineering and Applied Sciences, Harvard University, Cambridge, Massachusetts}
\end{center}

\newpage
\begin{abstract}
\cite{Munk1966} showed that the deep (1000-3000 m) vertical temperature profile is consistent with a one-dimensional vertical advection-diffusion balance, with a constant upwelling and an interior diapycnal diffusivity of $\mathcal{O}(10^{-4})$ m$^{2}$ s$^{-1}$. However, typical observed diffusivities in the interior are $\mathcal{O}(10^{-5})$ m$^{2}$ s$^{-1}$. Recent work suggested that the deep stratification is set by Southern Ocean (SO) isopycnal slopes, fixed by SO eddies, that communicate the surface outcrop positions to the deep ocean. It is shown here, using an idealized ocean general circulation model, that SO eddies alone cannot lead to the observed exponential temperature profile, and that interior mixing must contribute. Strong diapycnal mixing concentrated near the ocean boundaries is shown to be balanced locally by upwelling. A one-dimensional Munk-like balance in these boundary mixing areas, although with much larger mixing and upwelling, leads to an exponential deep temperature stratification, which propagates via isopycnal mixing to the ocean interior. The exponential profile is robust to vertical variations in the vertical velocity, and persists despite the observed weak interior diapycnal mixing. Southern Ocean eddies link the surface water mass transformation by air-sea fluxes with the deep stratification, but the eddies do not determine the stratification itself. These results reconcile the observed exponential interior deep temperature stratification, the weak diapycnal diffusivity observed in tracer release experiments, and the role of Southern Ocean dynamics.
\end{abstract}


\section{Introduction}
\label{sec:introduction}

\begin{figure}
  \begin{center}
    \includegraphics[width=1.0\textwidth,angle=0]{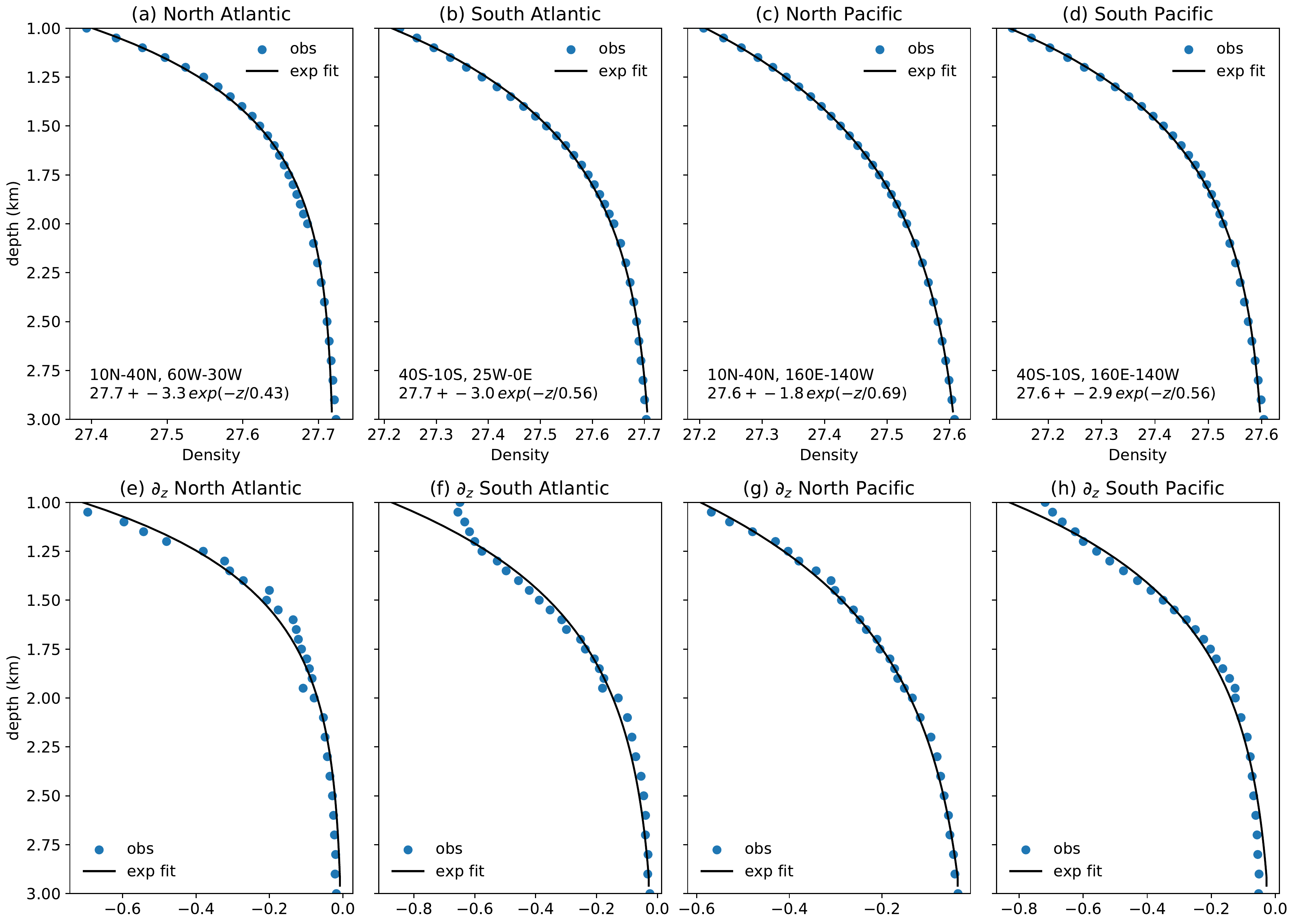}\quad
  \end{center}
  \caption{The observed interior mid-depth potential density and corresponding exponential fits, for the North and South Atlantic and Pacific oceans (upper panels), and the vertical derivatives of the potential density (lower panels, multiplied by 1000).}
  \label{fig:obs-middepth-temp}
\end{figure}

The observed deep, 1-3 km depth, ocean interior vertical potential density profile and its derivative, can both be fit to very good accuracy by an exponential function in many ocean regions, as shown by the four examples of Fig.~\ref{fig:obs-middepth-temp}. As noted by \cite{Munk1966}, an exponential temperature profile is also the solution to the one-dimensional vertical advective-diffusive balance, $w{\partial T}/{\partial z} = \kappa_v {\partial^2 T}/{\partial z^2}$ with a constant diffusive mixing coefficient $\kappa_v=10^{-4}$ m$^2$ s$^{-1}$ and a constant upwelling $w=10^{-5}$ m s$^{-1}$. While global water mass budget calculations are consistent with such mixing values \cite[][]{Lumpkin2007}, direct measurements of $\kappa_v$ in the ocean interior based on turbulent dissipation measurements and tracer release experiments suggest ocean interior vertical mixing values that are ten times smaller, $\mathcal{O}(10^{-5})$ m$^{2}$ s$^{-1}$ \cite[][]{Polzin1997, Ledwell1998}, leaving the cause of the exponential stratification unexplained.

Extending the \cite{Munk1966} idea and connecting it with adiabatic thermocline theories that require specification of the shallow thermocline stratification \cite[][]{Luyten-Pedlosky-Stommel-1983:ventilated, Rhines-Young-1982:theory}, \cite{Tziperman-1986:role} suggested that both the deep and shallow stratification are set by a balance between the net water mass formation near the surface into a given range of isopycnals, and the cross-isopycnal fluxes due to interior mixing. More specifically, cooling at the surface of the SO, for example, may lead to a net water-mass formation within some isopycnal range, which must be balanced at a steady state by the net cross isopycnal mass flux into this same range in the ocean interior. The interior fluxes are driven by small-scale mixing which depends on the vertical stratification there. If the surface formation and interior mixing effects do not balance, the outcrop position and resulting air sea fluxes, as well as the deep stratification, will adjust until such a balance is obtained. \cite{Tziperman-1986:role} further showed that a Munk-like vertical balance produces a very-nearly exponential temperature stratification even when the upwelling and diffusivity are not constant in the vertical ($z$) direction. \cite{Samelson-1998:large}, while not addressing the deep exponential profile explicitly, suggested that the discrepancy between the large-scale deep stratification and the weak observed vertical mixing may be explained by strong localized mixing near ocean boundaries. \cite{Marotzke-1997:boundary} made a related point, demonstrating the effects of boundary mixing on the overturning at a very coarse model resolution. In fact, as Marotzke notes, \cite{Munk1966} already suggested a significant role for strong mixing along the ocean boundaries.

Recent studies have proposed an alternative explanation for the deep ocean stratification based on a significant role for Southern Ocean (SO) eddies rather than deep vertical mixing. The general idea discussed in these works is that isopycnals in the southern ocean connect the surface outcrops with the deep stratification north of the SO, and that these sloping SO isopycnals are set by eddy processes as well as drive a residual circulation there. \cite{Marshall-Radko-2006:model} analyze the "upper-cell" of the SO (down to two km). They prescribe the vertical stratification of the interior north of the SO, and the surface buoyancy and use these boundary conditions to calculate the eddy-driven circulation. \cite{Ito-Marshall-2008:control} consider the SO at deeper levels, where mixing may be important, and again do not attempt to explain the interior stratification away from the SO and its exponential character. \cite{Wolfe2010} argued that the mid-depth stratification, 1-3 km depth, is set primarily by the isopycnal slopes in the Antarctic Circumpolar Current (ACC), and that the stratification is therefore independent of the value of $\kappa_v$. \cite{Nikurashin2011,Nikurashin2012} generalized the treatment of previous work regarding the role of SO eddies, and considered both high and low vertical mixing limits. They suggest that diapycnal mixing is not necessary to maintain a deep stratification, while the large mixing limit may be of relevance for the abyssal ocean. The state, for example, ``In the limit of weak diapycnal mixing, typical for the middepth ocean, deep stratification throughout the ocean is produced by the effects of wind and eddies in a circumpolar channel and maintained even in the limit of vanishing diapycnal diffusivity and in a flat-bottomed ocean.'' \cite{Shakespeare-McCHogg-2012:analytical} also study the response of the ocean to SO winds and prescribed surface buoyancy fluxes, yet their 3-layer formulation does not allow them to examine the deep exponential stratification profile. Finally, \cite{Mashayek2015} and \cite{Ferrari2016} also argue that the Southern Ocean is playing a dominant role and add the contribution of large diapycnal mixing due to a rough bottom topography.

Some of the above works do not show a solution for the vertical interior density profile that allows to examine its exponential character. The deep stratification predicted by \cite{Wolfe2010} and \cite{Nikurashin2011} at low values of the interior diapycnal mixing, which they consider the realistic mid-depth parameter regime, are both decidedly non-exponential as further discussed in our analysis below. More generally, none of the above works on the role of the SO addresses the exponential shape of the deep stratification, and they therefore do not attempt to explain this dominant feature of the observations.

The present paper focuses on the exponential mid-depth stratification (1-3 km, also referred to here as deep stratification, and always excluding the upper and lower 1 km). We examine the connections between the primary processes that have been proposed: interior diapycnal mixing, boundary mixing, and SO eddy dynamics and isopycnal slopes. We argue, in contrast to the above studies, that the eddy dynamics setting the SO isopycnal slopes cannot be responsible for the observed exponential deep stratification at weak to vanishing values of the diapycnal mixing. Instead, we show that enhanced deep boundary mixing, interacting with surface water mass transformation, leads to an exponential profile equivalent to the Munk picture even when the interior diapycnal mixing is very weak. SO eddy dynamics and isopycnal slopes play a critical role in communicating between the surface and deep water mass transformation processes, but do not set the exponential stratification. In order to focus on only the relevant factors, we employ idealized numerical experiments in a basin-channel configuration of an ocean model similar to \cite{Wolfe2010}. We prescribe enhanced mixing near the horizontal ocean boundaries similar to \cite{Marotzke-1997:boundary} and \cite{Samelson-1998:large}, but we additionally include a SO-like channel which they have not, and discuss the exponential stratification which was not addressed by these studies. Previous idealized studies \cite[e.g.,][]{Nikurashin2011, Shakespeare-McCHogg-2012:analytical} considered horizontally integrated diapycnal mixing throughout the ocean basin, and their results are thus not inconsistent with those due to spatially inhomogeneous mixing explored here more explicitly. However, there is value in examining the explicit role and effect of such inhomogeneous mixing, as done here.

The mid-depth ocean stratification deviates from exponential in some regions, while in others the density is very nearly exponential although the temperature is not (e.g., within the South Atlantic ocean north of the Southern Ocean). While a more thorough quantification of where the stratification is exponential is outside the scope of this study, the examples shown in Fig.~\ref{fig:obs-middepth-temp} are sufficiently intriguing and representative of large areas that they justify the discussion attempted here. The exponential depth scales shown in Fig.~\ref{fig:obs-middepth-temp} are generally less than the canonical 1 km value calculated by Munk. It is difficult to directly deduce a vertical mixing estimate from this scale, as that scale also depends on the strength of upwelling. Our objective is to form a consistent \textit{qualitative} picture that incorporates an exponential scale with very weak ocean interior mixing. We therefore proceed in the followings to attempt to reconcile this canonical value with the observed low interior diapycnal diffusivity, leaving it to future studies to examine the difference between different ocean basins and to re-evaluate Munk's recipes based on updated values for both the upwelling and diffusivities.

Observational evidence, theory and modeling all indicate elevated diapycnal diffusivity near rough topography, ocean boundaries and ocean passages \cite[e.g.,][]{Polzin1997, Thurnherr2011, StLaurent2007, Polzin1997, StLaurent2012, Sheen2013, Waterhouse2014, Nash2012, Melet2013}. While observations of mixing rates and turbulent diffusivity are sparse, they indicate that much of the missing diapycnal mixing may be found at the ocean margins. Currently available direct observations of small-scale diapycnal mixing \cite[e.g.,][]{Waterhouse2014} do not allow determining with certainty if the small areas of strong mixing are equivalent, in an integral sense, to a homogeneous Munk-like vertical diffusivity everywhere. 

In this study we examine the possible role of such isolated mixing and its interaction with the SO dynamics, in order to develop a conceptual understanding of a possible scenario that sets the mid-depth exponential stratification profile. We do not attempt a realistic simulation, developing instead an idealization that demonstrates the physics that determines the exponential deep stratification. One possible complication not addressed here is that the deep ocean under the main thermocline includes both isopycnals that outcrop in the SO and those that outcrop in the North Atlantic, leading to possibly different dynamics \cite[][]{Wolfe2010}.

\section{Ocean Model and experiments}
\label{sec:ocean-model}

We show results from three numerical experiments in an idealized ocean basin with an Antarctic Circumpolar Current (ACC) channel. Our experimental design combines experiments similar to those of \cite{Samelson-1998:large} with a configuration that contains a SO-like channel similar to \cite{Wolfe2010}. We use the Massachusetts Institute of Technology generalized circulation model (MITgcm) hydrostatic ocean model \cite[][]{Marshall1997}. The domain is a 3500 m deep, flat-bottomed rectangular box spanning 60 degrees in longitude and 140 degrees in latitude (70$^{\circ}$S to 70$^{\circ}$N) at a 1$^{\circ}$ horizontal resolution. In the Southern Hemisphere, there is a zonally re-entrant channel that spans 70$^{\circ}$S to 50$^{\circ}$S. There are 45 vertical levels ranging from a thickness of 10 m in the surface layer to 261.5 m in the lowest layer. The model equations are solved on a spherical polar grid. Sub-gridscale mixing is parameterized with Gent-McWilliams/ Redi isopycnal mixing \cite[][]{Gent1995, Redi1982} and a K-Profile Parameterization of vertical mixing \cite[][]{Large1994}. Background Laplacian and biharmonic horizontal viscosities are $10^{6}$ m$^2$ s$^{-1}$ and $10^{10}$ m$^{4}$ s$^{-1}$ for eliminating grid-scale noise. Background vertical viscosity is set to $10^{-3}$ m$^2$ s$^{-1}$. The salinity is set to a uniform value, and the equation of state is linear in temperature with $\rho_0=1028.665$ kg m$^{-3}$, $T_0=20^\circ$C and $\alpha=1665.22\times 10^{-7}$ kg m$^{-3}$ $^\circ$C$^{-1}$.

\begin{figure}
\centering\includegraphics[width=0.5\textwidth,angle=0]{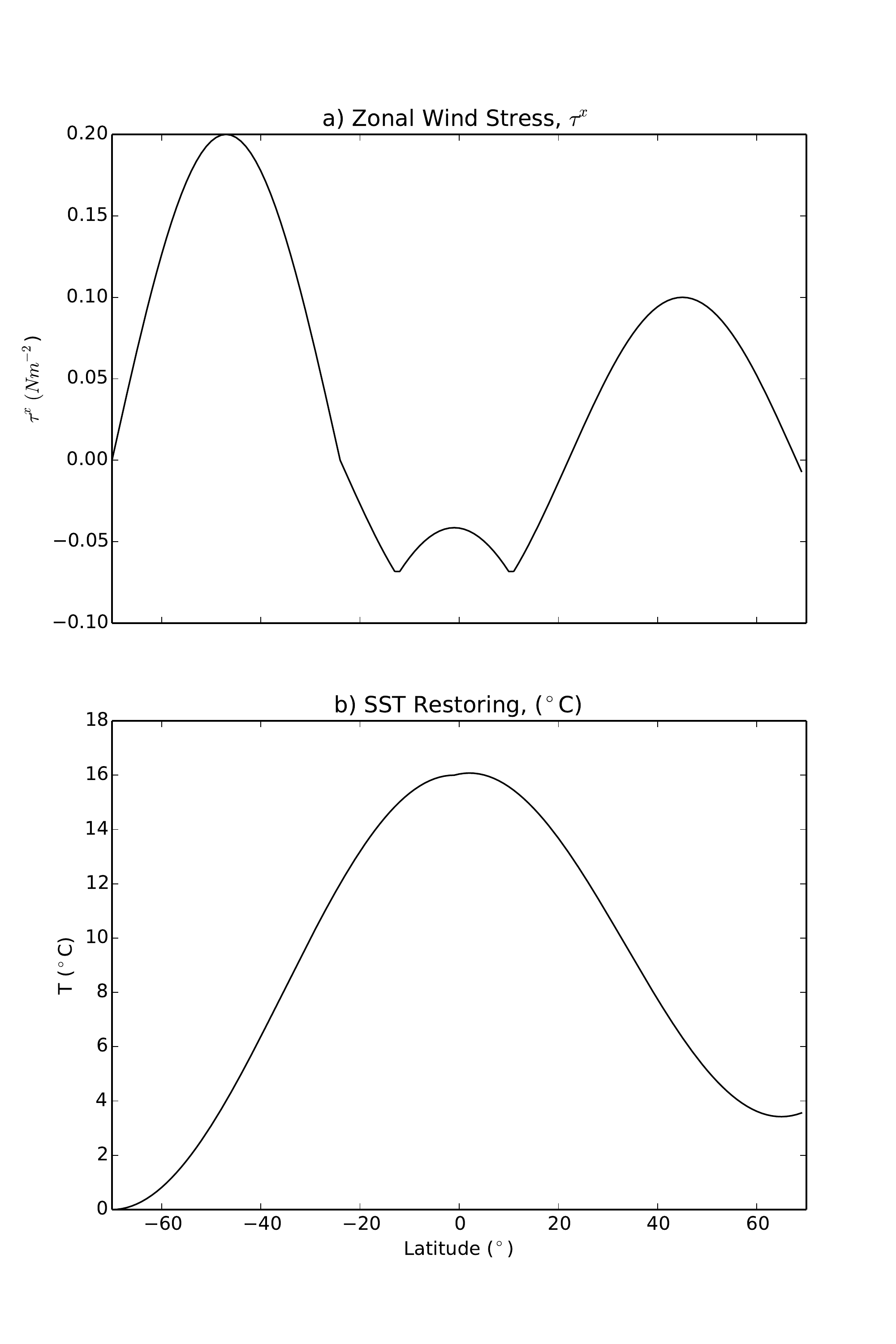}
\caption{Surface forcing fields for all experiments: a) Zonal wind stress b) SST.}
\label{fig:surface-forcing}
\end{figure}

The surface forcing, shown in Fig.~\ref{fig:surface-forcing} is modeled after \cite{Wolfe2010} to be both idealized and generically representative of modern meridional asymmetry in surface wind and temperature fields in the Pacific Ocean. The zonal surface wind is zonally symmetric and the wind maximum over the re-entrant channel is 0.2 N m$^{-2}$, twice the maximum of 0.1 N m$^{-2}$ in the Northern Hemisphere. There are also two relative minima in wind stress of $-0.07$ N m$^{-2}$ bounding the equator. There is no meridional wind forcing. Temperature in the surface layer is relaxed towards a zonally-symmetric temperature field on a 1-week timescale to the values shown in Fig.~\ref{fig:surface-forcing}b. As in the observed Pacific Ocean, the restoring surface temperature in the Southern Hemisphere is colder than the surface temperature in the Northern Hemisphere. We accelerate the model experiments to steady state using asynchronous time-stepping \cite[][]{Bryan1984} with deltaTmom=300s and deltaTtracer=3000s for 3500 model years. We then confirm the model is at steady state by running it for 100 years with synchronous time-steps (deltaTmom=deltaTtracer=300s) and take the steady-state values as when the trends of SST, $\Theta$ and KE are are less than 1$\%$ of their respective 20-year mean values. All steady-state quantities are averaged over a 10-year time interval integrated with synchronous time steps. However, the integrations approach a steady state, as the model is of a coarse resolution and shows no variability at steady state. 

\begin{figure*}
\centering\includegraphics[width=0.7\textwidth,angle=0]{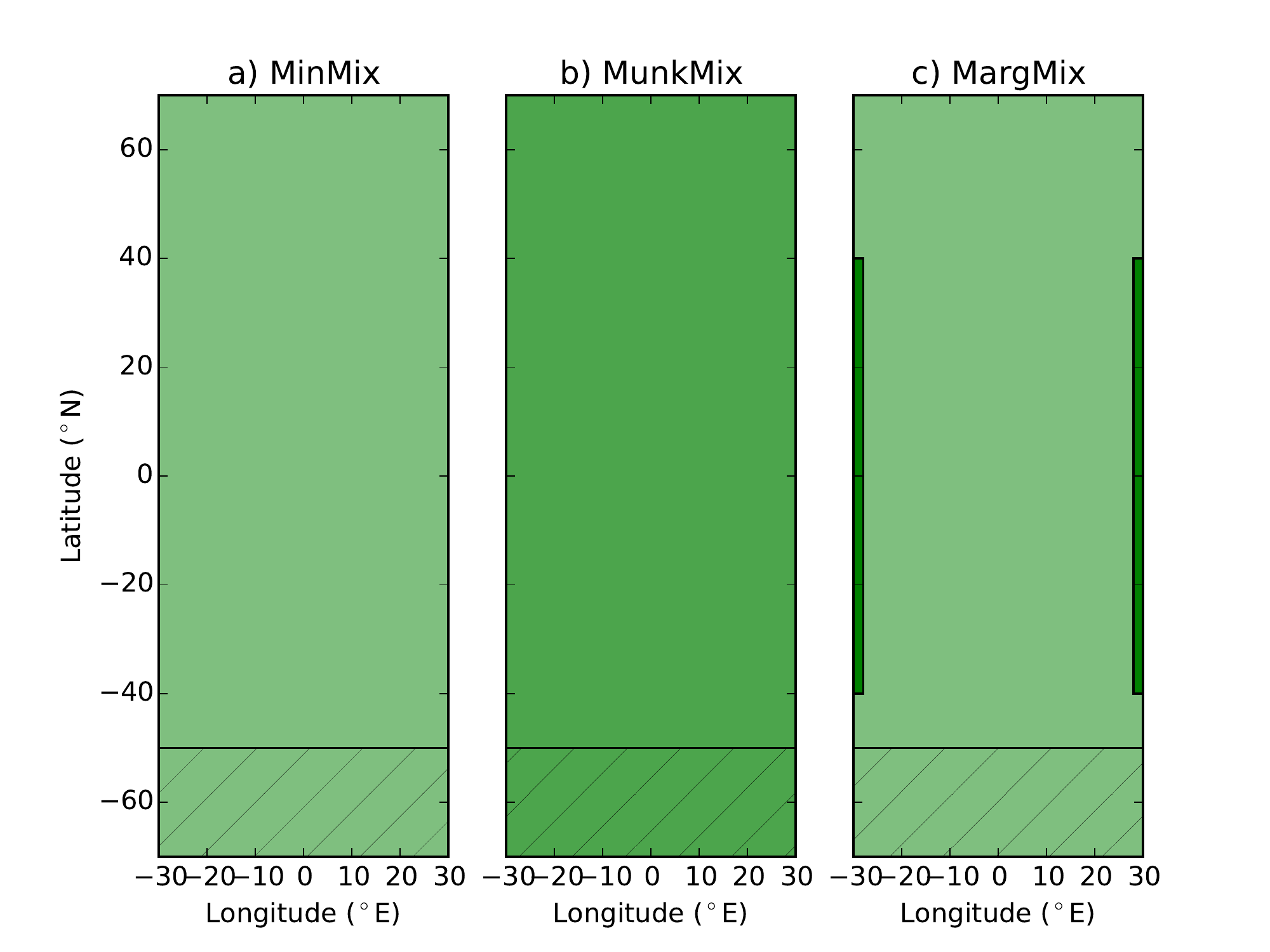}
\caption{Schematic of variation in background vertical diapycnal diffusivity for all experiments. Re-entrant channel is indicated by hatch marks between 70S and 50S. a) MinMix: homogeneous low vertical mixing of $10^{-5}$ m$^2$ s$^{-1}$. b) MunkMix: homogeneous higher mixing, $10^{-4}$ m$^2$ s$^{-1}$. c) MargMix: low interior mixing, $10^{-5}$ m$^2$ s$^{-1}$, with elevated values of 3$\times 10^{-3}$ m$^2$ s$^{-1}$ indicated by dark green rectangles near eastern and western boundaries.}
\label{fig:kappa}
\end{figure*}

The experiment termed MinMix has a spatially constant vertical temperature diffusivity equal to $10^{-5}$ m$^2$s$^{-1}$, motivated by ocean interior observations. A second experiment, motivated by Munk's Abyssal Recipes and termed MunkMix, has $10^{-4}$ m$^2$ s$^{-1}$ vertical diffusivity. In the third experiment, MargMix, following \cite{Marotzke-1997:boundary} and \cite{Samelson-1998:large}, $\kappa_v$ is $10^{-5}$ m$^2$ s$^{-1}$ in the interior of the domain and 3$\times 10^{-3}$ m$^2$ s$^{-1}$ in the western and eastern boundary margins, which are each 2 degrees in longitude and 80 degrees in latitude (Fig.~\ref{fig:kappa}). The transition from high boundary mixing to background mixing is abrupt rather than gradual \cite[][]{Samelson-1998:large, Marotzke-1997:boundary}, yet a close examination of the solution shows no resulting numerical noise. The vertical diffusivity at the margins in MargMix is chosen to produce an area-average diffusivity similar to that of the MunkMix experiment. Numerical implicit diapycnal mixing is non-negligible, but its effects on stratification in our experiments are equivalent to an explicit vertical diffusivity of less than $\mathcal{O}(10^{-5})$ m$^2$ s$^{-1}$, as described in the results section below.

For averaging purposes we define the ocean interior as the area between 40$^\circ$S and 40$^\circ$N. Interior zonal averages for MargMix also exclude the boundary region in which diapycnal diffusivity is elevated. The zonally-integrated meridional overturning stream function, $\psi$, is defined as,
\begin{equation}
  \psi(\theta,z) = - \int_{\phi_W}^{\phi_E} \int_{-H}^z v \, r\cos\theta dz d\phi.
\end{equation}
where $\phi$ is longitude, $\theta$ is latitude, $H$ is the bottom depth, $z$ is the vertical coordinate, $r$ is the radius of the earth and $v$ is the meridional velocity. 
Given the vertical velocity $w(z)$ and the diffusivity $\kappa_v$ at a specific longitude and latitude, we calculate a Munk-like prediction of the temperature profile at a given horizontal location, $T_{Munk}(z)$, that solves $wT_z=\kappa_vT_{zz}$.  The solution is given by,
\begin{align}
  \label{eq:Munk}
  T_{Munk}(z)=C_1\int_{-H}^z \exp\left(\int_{-H}^{z'}
  \frac{w(z'')}{\kappa_v}\,dz''\right)\,dz' + C_2,
\end{align}
where $H=3000$ m (not the bottom depth of 3500 m) and the solutions are calculated only between 1000-3000 m depth. An optimization is used to find the integration constants $C_1$ and $C_2$ such that the solution is the best least-square fit to the model interior temperature at this location. The vertical profile $T_{Munk}(z)$ calculated separately at each horizontal location ($\phi$, $\theta$), is then averaged horizontally. This technique is not equivalent to calculating $T_{Munk}(z)$ using the ocean-average value of $w$ due to nonlinear term in the equation for $T_{Munk}$ involving the spatially variable $w$.

\section{Results}
\label{sec:results}

\begin{figure*}
\centering\includegraphics[width=\textwidth,angle=0]{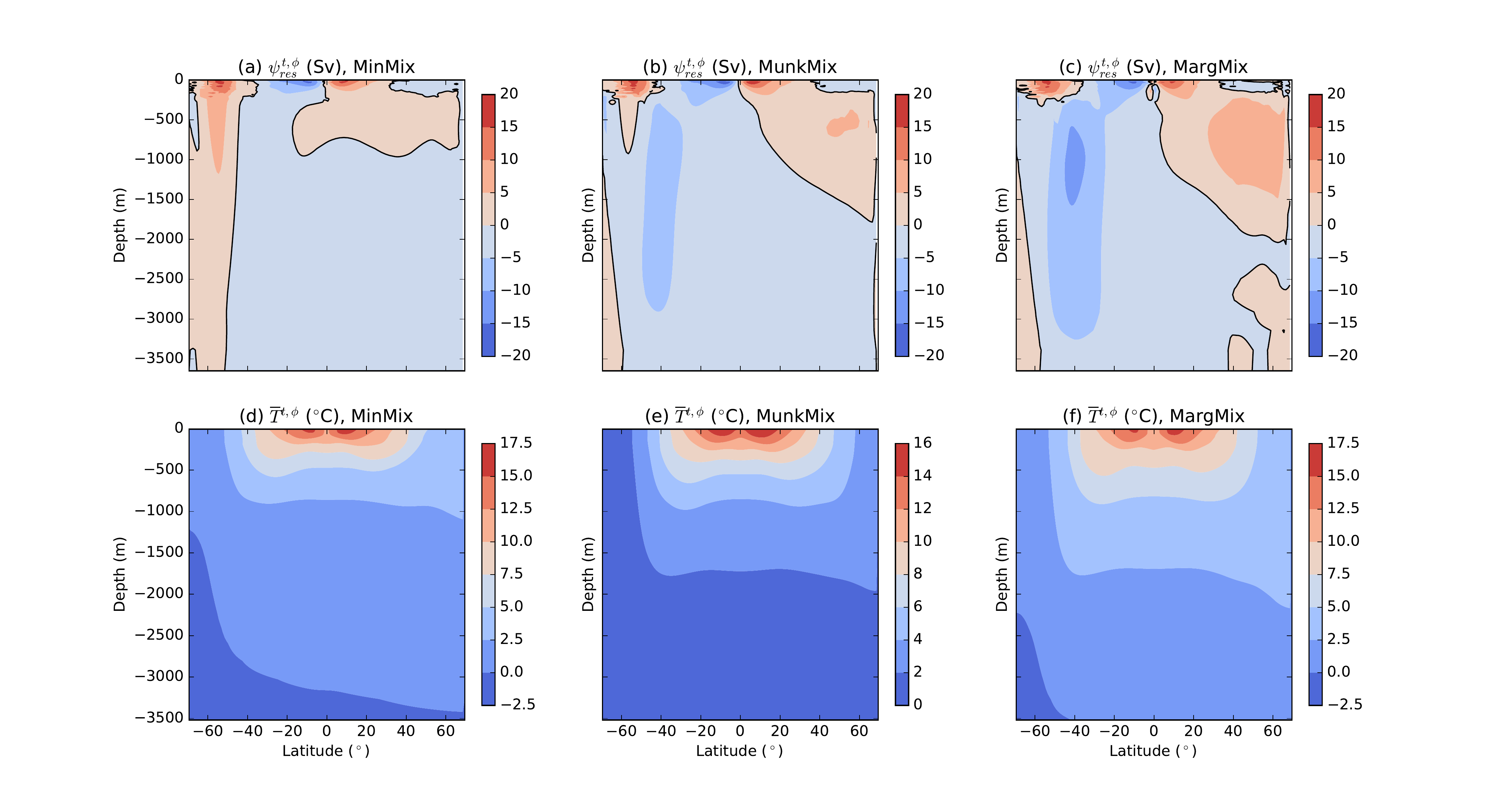}
\caption{Steady-state MOC residual stream function $\psi_{res}$, and area-weighted zonal mean $T$, excluding West and East boundaries, for (a,d) MinMix case (b,e) MunkMix case (c,f) MargMix case. MOC reflects Eulerian velocities plus bolus velocities calculated by the GM-Redi parameterization.}
\label{fig:zonal-means}
\end{figure*}

The large-scale horizontal circulation and temperature distribution in all three experiments are qualitatively similar to observed distributions (Fig.~\ref{fig:zonal-means}). The zonally-averaged temperature has similar spatial structure in all cases. The surface and intermediate isopycnals outcrop in both hemispheres, and the deepest interior isopycnals outcrop in the Southern Hemisphere but not in the Northern Hemisphere. The position of the isopycnal outcrops does not vary significantly between the experiments, but the intermediate-depth stratification, which is our focus, does, as shown by the density of the isotherms in Fig.~\ref{fig:zonal-means}(d--f) away from the poles. The weakest interior stratification is found in the MinMix experiment, while the stratification in MunkMix and MargMix are similar.

The zonally-averaged meridional residual stream function \cite[][]{Marshall-Radko-2003:residual, Wolfe2010}, $\psi_{res}$, is strongest in the MargMix experiment and weakest in the MinMax one \cite[][]{Bryan-1987:parameter, Marotzke-1997:boundary, Samelson-1998:large}. There are clearly defined North Atlantic and Southern Ocean overturning cells in MunkMix and MargMix. The eddy-driven SO cell is opposite in sign to the Eulerian one (not shown), leading to a partial cancellation except near the surface in the SO where the parameterization of the eddy-driven circulation breaks down due to the vertical isopycnals. The stratification differences of interest to us here appear in the zonal averages below 1000 m.

\begin{figure*}
  \centering\includegraphics[width=\textwidth,angle=0]{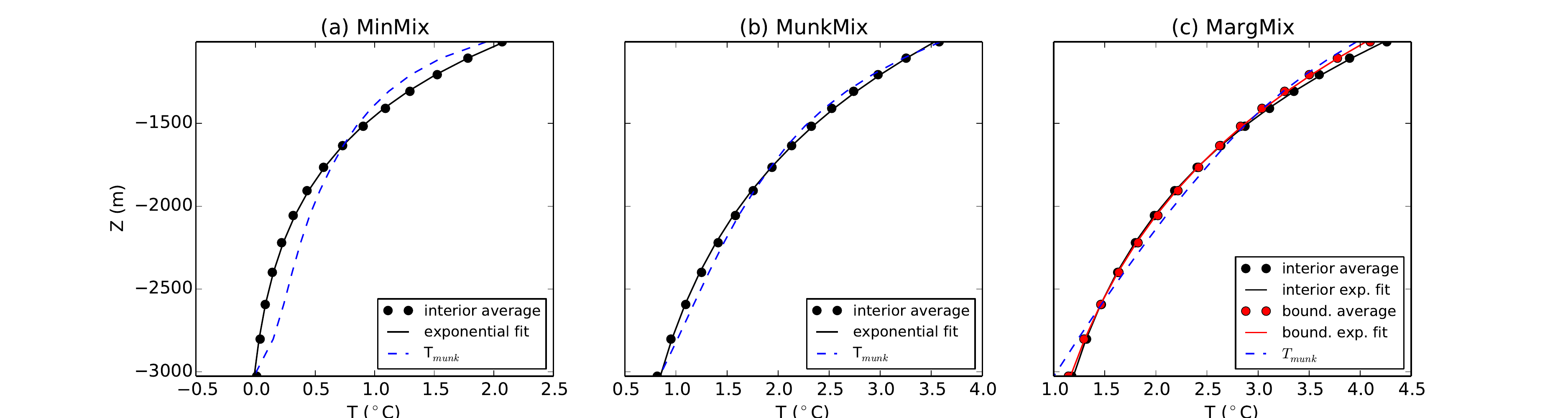}
 \caption{Steady-state horizontally averaged temperature solution $T(z)$,  $T_{Munk}(z)$ solution using eqn (\ref{eq:Munk}), and an exponential fit for: (a) Interior average (see section \ref{sec:ocean-model}) of MinMix solution; (b) interior average of MunkMix solution. (c) MargMix case: both interior and boundary averages calculated within 4 degrees longitude of the West and East boundaries. $T_{Munk}(z)$ is calculated and averaged in the interior for MinMix and MunkMix and averaged in the boundary region (both W and E) for the MargMix experiment. Within the respective regions, the standard deviation of $T_{Munk}(z)$ is 0.1-0.3 $^\circ$C.}
 \label{fig:temp-profiles}

\end{figure*}

Despite large differences in the spatial structure of the diapycnal diffusivity, the interior temperature profiles for MunkMix and MargMix are almost identical, as also found by \cite{Samelson-1998:large}. Our focus, though, is specifically the existence and mechanisms leading to an exponential profile, which he did not address. We average the intermediate-depth temperature profiles over the interior domain (40$^{\circ}$S to 40$^{\circ}$N and 4$^{\circ}$ inward from the eastern and western boundaries, and shown over a depth range of 1000 to 3000 m) and find the least-squares fit to exponential functions of the form $A + Be^{-z/h}$, where $A,B$ and $h$ are constants (Fig.~\ref{fig:temp-profiles}). The vertical decay scales, $h$, for the MinMix, MunkMix and MargMix cases are, respectively, 654, 1082 and 1042 m. The MunkMix and MargMix interior mid-depth stratification profiles are not only exponential but also have nearly identical vertical decay scales to each other and to the Pacific profile fitted by Munk, which had a roughly 1km decay scale. Most importantly, all three experiments are driven by the same SO winds, yet still show differences in the deep stratification, indicating that the SO does not set the deep stratification by itself.

\begin{figure*}
  \centering\includegraphics[width=\textwidth,angle=0]{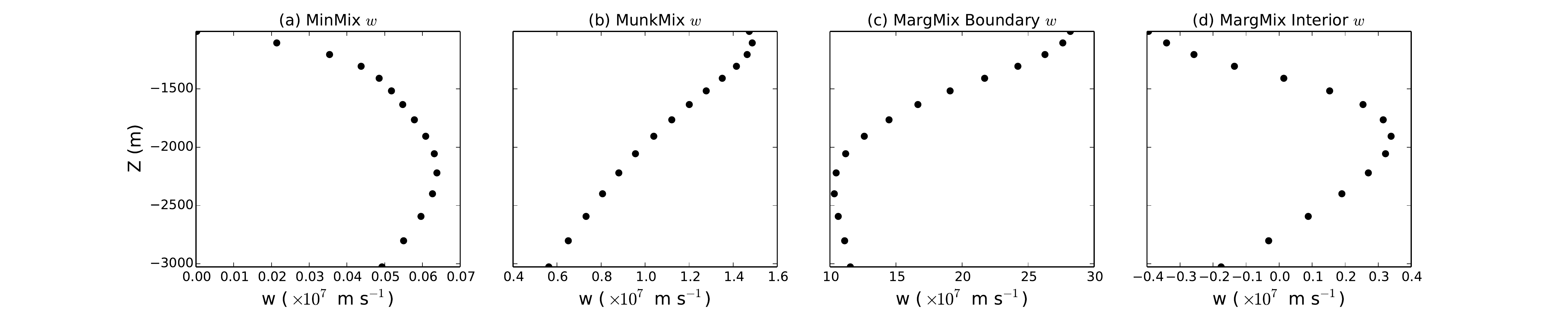}
  \caption{Steady-state horizontally averaged $w(z)$ used to calculate $T_{Munk}$: (a) Interior average of MinMix solution; (b) Interior average of MunkMix solution; (c) MargMix case: averaged over boundary areas.}

  \label{fig:w-profiles}
\end{figure*}

As shown in Fig. 5, the MargMix average temperature profile in the interior (where $\kappa_v=10^{-5}$ m$^2$ s$^{-1}$) is essentially identical to the MargMix boundary average temperature profile (where $\kappa_v=3\times10^{-3}$ m$^2$ s$^{-1}$). The boundary profile in this experiment is determined by a balance between the large upwelling and large mixing in the boundary areas, as demonstrated by the profile of $T_{Munk}(z)$ computed for this experiment using Eq.~(\ref{eq:Munk}) (dashed lines in Fig.~\ref{fig:temp-profiles}). The vertical velocity in the boundary regions of MargMix is ${\cal O}(10^{-5})$ m/s, much larger than the interior values in all three experiments (Fig.~\ref{fig:w-profiles}).

We note that while the stratification in both MargMix and MunkMix is exponential, the vertical velocity profiles $w(z)$ are not constant in the vertical. The exponential shape is thus very robust, in spite of this deviation from the Munk hypothesis, as noted by \cite{Tziperman-1986:role}. The exponential boundary vertical temperature profile is communicated to the interior via horizontal/isopycnal mixing, consistent with the suggestion of \cite{Munk1966}, and as also found by \cite{Samelson-1998:large}. A passive tracer experiment in the MargMix experiment (not shown) shows that a tracer released in the interior, is mixed horizontally to the boundary, mixed vertically within the boundary region, and then mixed back into the interior at different vertical levels, consistent with the view of boundary mixing acting as effective vertical mixing for the interior as well.

The Munk balance solution (Eq.~\ref{eq:Munk}) for the temperature profile in MinMix is a poorer fit to the model temperature profile than it is in the MunkMix or MargMix experiments (compare dashed and solid lines in Fig.~\ref{fig:temp-profiles}a, vs Fig.~\ref{fig:temp-profiles}b,c) because in MinMix the implicit numerical diffusivity is not negligible relative to the explicit diffusivity, $\mathcal{O} (10^{-5})$ m$^2$ s$^{-1}$. An experiment with zero explicit vertical diffusivity yields exponential stratification with a 386 m depth, comparable to the 654 m depth scale in MinMix. Thus, because the non-negligible implicit numerical diffusivity is not accounted for in the value of $\kappa_v$ used in the computation of the Munk solution, the temperature profile obtained from Eq. (\ref{eq:Munk}) is slightly biased. This effect is even less significant in MunkMix and MargMix because the explicit vertical diffusivities are at least an order of magnitude greater than the implicit numerical diffusivity.

We also note that while the averaged vertical velocity is positive in the concentrated mixing region of MargMix as expected, it is negative in the interior of MargMix between 2500 and 3000 m (Fig.~\ref{fig:w-profiles}d). At the MargMix boundaries, the average vertical velocity is not only positive, but also three orders of magnitude greater than the MargMix interior-average vertical velocity. The sinking due to surface cooling at high latitudes is therefore balanced by lower-latitude upwelling in the margins, and the much smaller interior velocity and vertical mixing are immaterial for setting the stratification and overturning circulation in MargMix \cite[see also][]{Marotzke-1997:boundary, Scott-Marotzke-2002:location}.

\section{Discussion}
\label{sec:discussion}

Recent studies suggested that the mid-depth stratification is set within the ACC, where the interplay between wind-driven Ekman transport and the balancing eddy-driven isopycnal relaxation controls the slopes of the isopycnals connecting high latitude surface of the Southern Ocean to the mid-latitude ocean interior. In this view the isopycnal locations are set at the surface by air-sea fluxes and communicated to the deep ocean by the above interplay \cite[][]{Wolfe2010, Nikurashin2011, Nikurashin2012, Mashayek2015, Ferrari2016}. However, the density and buoyancy frequency profiles shown in some of these papers are not exponential. One possible mechanism for setting an exponential profile via SO processes (Raffaele Ferrari, personal communication) is that if planetary potential vorticity (PV), $f\rho_z$, and density $\rho$ are both conserved on the SO interior isopycnals, the two must have a functional relationship. Strictly this is the case along a streamline, unless eddies mix PV along isopycnals, but we persist and assume for a moment that such a relation applies throughout the SO, and further assume for simplicity that it is a linear functional relation, $f\rho_z=\lambda\rho$, which leads to an exponential profile $\rho\sim\exp(\lambda{z}/f)$.

\begin{figure*}
  \centering\includegraphics[width=\textwidth,angle=0]{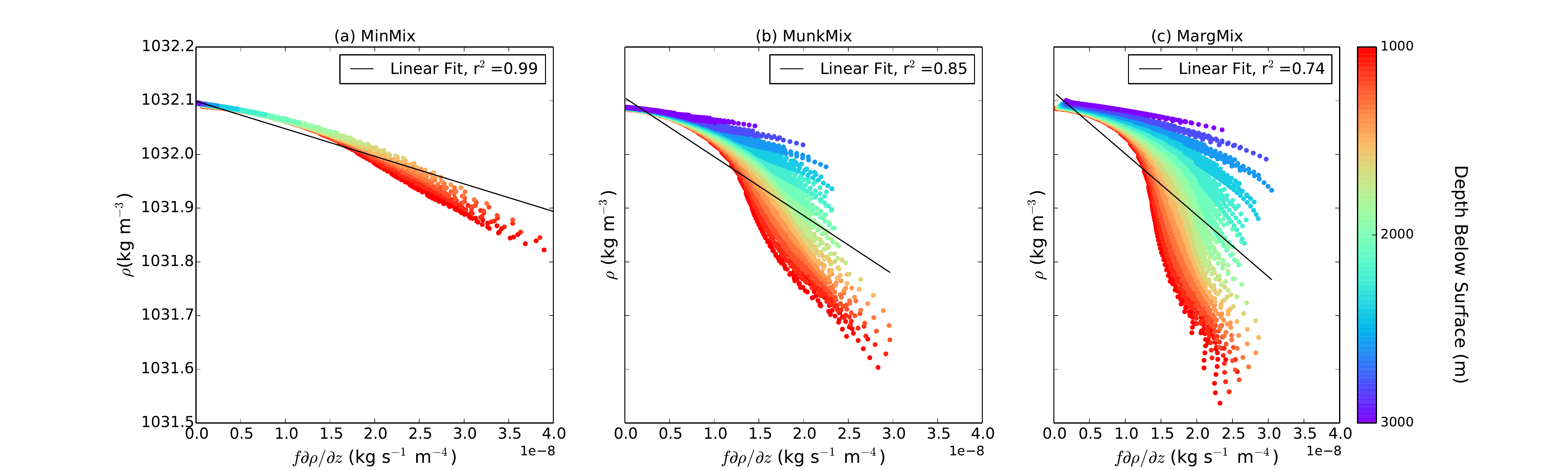}
  \caption{A scatter plot of planetary PV $f\rho_z$ vs density $\rho$ in the southern ocean between 1 and 3 km depth, with colors indicating the depth from which a given point on the scatter plot is taken. The black line on each plot is a linear regression of the values at all depths. (a) MinMix (b) MunkMix (c) MargMix.}
  \label{fig:pv-scatter}
\end{figure*}

While the above assumptions used to derive this exponential profile may not be applicable in our model, we demonstrate how this linear functional relationship between planetary potential vorticity and density may be tested by examining scatter diagrams of $\rho$ vs $f\rho_z$ for the different experiments (Fig.~\ref{fig:pv-scatter}) and calculating a linear fit to all of the data between 1000 and 3000 m depth. The results do not support a linear relation (or any relation) on a single isopycnal, as they show many values of PV for each value of $\rho$, except near $\rho_z = 0$. MunkMix and MargMix experiments in particular, which have high interior diapycnal diffusivity, deviate significantly from PV conservation on isopycnals. The slopes of the calculated linear fits would be consistent with the exponential fit for the mid-depth density profiles examined above only within an order of magnitude. Due to this large mismatch, we conclude that the parameterized SO eddies in our model runs do not determine the temperature interior mid-depth stratification depth scale.

Rather, our results suggest if a clear functional relation between the SO PV and density is found in observations or in an eddy-resolving model, it could be determined by the interplay between deep vertical mixing in the mid-depth interior of the ocean and the surface water mass transformation, as follows. While the outcrop locations in the Southern Ocean and the deep stratification are connected by isopycnal slopes in the Southern Ocean that are set by eddies, the position of the outcrops and the depth of the deep isopycnals, require additional physics to be determined, such as the condition that the net mass flux across any isopycnal surface must vanish at a steady state \cite[][]{Tziperman-1986:role}. This balance includes cross-isopycnal fluxes at the surface of the SO due to water mass transformation driven by air-sea fluxes \cite[][]{Walin1982, Tziperman-1986:role, Speer-Tziperman-1992:rates}, and cross-isopycnal fluxes north of the SO, driven by small-scale diapycnal mixing in the boundary mixing regions. The outcrop locations of isotherms in the SO affect the air-sea heat fluxes and thus the near-surface cross-isothermal (and cross-isopycnal) water mass transformation. These outcrop locations and the water mass volume injected by air-sea fluxes between any two isopycnals are then communicated to the deep ocean via the sloping isopycnals in the ACC, affected by eddies. In the deep ocean, the water mass transformation due to upwelling across an isopycnal surface is driven by diapycnal mixing. If the net mass flux into the space between two isopycnals, due to surface and deep transformations, do not exactly balance, both the surface isopycnal positions in the SO and the deep stratification will adjust until such a balance is achieved and an equilibrium is reached. The adjusting stratification will affect and include the SO stratification as well, setting the above functional relationship.

The deep stratification set by this balance between surface and interior water mass transformation processes tends to be exponential in form because, as was pointed out by \cite[][]{Tziperman-1986:role} and confirmed here, while an exponential profile is an exact solution for constant upwelling and diffusivity \cite[][]{Munk1966}, the stratification obtained from $w(z)\rho_z=(\kappa(z)\rho_{z})_z$ is very nearly exponential \emph{even if the upwelling or diffusivity are not uniform.}

\begin{figure*}
  \centering\includegraphics[width=\textwidth,angle=0]{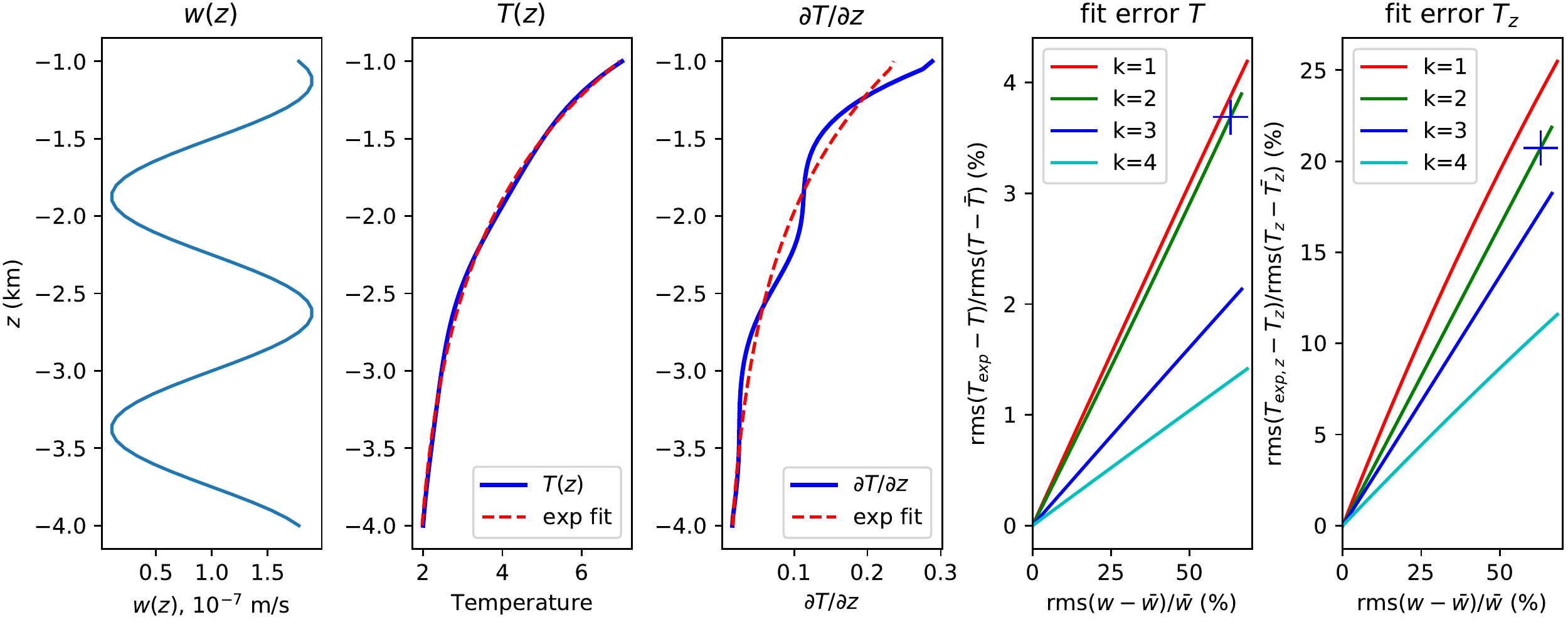}
  \caption{Demonstrating the insensitivity of the exponential profile to variations in the vertical velocity profile. (a) a sinusoidal vertical velocity profile with wavenumber $k$ = 4  (b) the corresponding temperature profile obtained from the Munk balance, and an exponential fit. (c) The numerical vertical derivative of the temperature profile and its exponential fit. (d,e) plots of the error in fit of the temperature and its vertical derivative to exponentials, as function of the amplitude of the vertical velocity variations from the mean ($w -\overline{w}$) and for different values of the wavenumber, $k$ of the imposed sinusoidal vertical velocity structure.}
  \label{fig:exponential-insensitivity}
\end{figure*}

To further demonstrate this insensitivity of the exponential profile, Fig.~\ref{fig:exponential-insensitivity}a-c shows a solution to the Munk balance for a strongly varying vertical velocity and its exponential fit. We define a fit quality measure for the temperature as rms$(T_\mathrm{exp}-T)/$rms$(T-\bar{T})$, where $T_\mathrm{exp}$ is the best exponential fit to the profile and $\bar{T}$ is the vertically averaged temperature. This is a nondimensional measure of the difference between the temperature and its fit, relative to the amplitude of the vertical variations in the temperature profile itself. A similar measure is defined for the vertical derivative of the temperature profile and its best exponential fit.

These fit quality measures are listed in Table~\ref{table:fit-error} for all relevant runs in this paper, plus some from relevant previous works. The vertical velocity is assumed sinusoidal in depth with a specified amplitude and vertical wavenumber. Fig.~\ref{fig:exponential-insensitivity}d,e show the quality of fit as function of the amplitude of vertical velocity variation, and for different wavenumbers. The amplitude of the vertical variations of the vertical velocity is measured again in nondimensional units, as rms$(w-\bar{w})/\bar{w}$. This figure provides justification for the above statement \cite[][]{Tziperman-1986:role} that the solution to the Munk balance tends to be nearly exponential even for a non-uniform vertical velocity profile. It is not surprising that the fit error is larger for the vertical derivative of the temperature, which is a more stringent measure of the quality of the exponential fit.

The error measures in the fit to an exponential profile based on the low internal diapycnal diffusivity results of \cite{Wolfe2010} and \cite{Nikurashin2012}, which they considered their realistic regime based on the measurements of diapycnal diffusivity in the ocean interior, are listed in Table~\ref{table:fit-error}, and show significant deviations from an exponential profile. The high diffusivity (CP-k8, $\kappa = 0.98 \times 10^{-4}$ m$^2$ s$^{-1}$) experiment from \cite{Wolfe2010} shows the best fit to an exponential profile, which is consistent with our MunkMix experiment results, but is of course not consistent with the observed low interior ocean diffusivity.

In the picture presented here, the eddies setting the SO isopycnal slopes play an important role in communicating between the deep mixing and surface transformation. All three processes must be playing a role in determining the exponential deep stratification.

\section{Conclusions}
\label{sec:conclusions}

We have examined the interaction between Southern Ocean dynamics and boundary-concentrated diapycnal mixing in the ocean interior, and demonstrated their role in setting the mid-depth stratification using idealized ocean model experiments. Our focus is specifically the very robust observed exponential profile originally identified by \cite{Munk1966}. In order to explain this exponential profile, we needed to deal with three challenges. First, the interior vertical diffusivity used by Munk to explain the exponential profile is much larger than that observed in the ocean interior in tracer release experiments. Second, the vertical velocity profile is not constant or spatially homogeneous as assumed by Munk. Third, recent studies have made physically convincing arguments that the deep stratification is primarily determined by SO dynamics: surface processes set the outcrop positions in the SO which are communicated to the deep ocean via SO isopycnal slopes set by eddy processes.

The first challenge is dealt with here following \cite{Samelson-1998:large, Marotzke-1997:boundary} and \cite{Scott-Marotzke-2002:location}, all of which considered the effects of boundary mixing but not the exponential stratification, by examining the possibility that concentrated vertical mixing near ocean boundaries can set the stratification which is then communicated to the ocean interior by near-horizontal isopycnal eddy mixing. We find that even in the presence of a SO-like channel, an exponential stratification cannot develop with a weak or vanishing diapycnal mixing. On the other hand, an enhanced vertical mixing near the ocean boundary, with small interior vertical mixing, can lead to a deep exponential stratification that is nearly identical to that found by having a uniform Munk-like vertical mixing throughout the ocean interior. In the concentrated mixing areas at the ocean margins there is a Munk-like balance of vertical advection and diffusion in our experiments. Both the vertical velocity and mixing at the margins are orders of magnitude larger than those assumed by Munk, but their ratio still yields an $\mathcal{O}(1000)$ m depth scale, similar to the observed interior profiles.

As for the second challenge, we demonstrated that the solution to the Munk balance of $w\rho_z=\kappa_v\rho_{zz}$ is very closely exponential even when the vertical velocity is not constant in the vertical direction, as was also shown by \cite{Tziperman-1986:role}. This robustness of the exponential profile was also demonstrated in the concentrated boundary mixing regions, where the vertical velocity is both very large and strongly varying in depth. Some previous works \cite[][]{Wolfe2010, Nikurashin2011, Nikurashin2012} showed that a mid-depth stratification on isopycnals that outcrop in both the North Atlantic and the SO may be determined by eddy processes without a role for cross-isopycnal mixing. However, we emphasize here that the stratification in that case is not expected to be exponential. The existence of an exponential profile at a depth as shallow as 1 km \cite[][]{Munk1966} together with our results here suggest that diapycnal mixing is likely a strong player even at this density range. The strong mixing limit of \cite{Nikurashin2011, Nikurashin2012} was suggested as being relevant to the abyssal ocean, and while these works did not address the exponential stratification profile, we add here that such a strong vertical mixing limit seems relevant in the deep ocean as well, via the action of boundary mixing, as demonstrated here.

These results are consistent with the hypothesis that the mid-depth stratification is determined by the condition that the net mass flux across any isopycnal surface must vanish at a steady state \cite[][]{Tziperman-1986:role}. Cooling at the surface of the SO near some isopycnal $\rho_1$, for example, leads to a water-mass transformation across this isopycnal and toward higher densities \cite[][]{Walin1982, Tziperman-1986:role, Speer-Tziperman-1992:rates}. This flux must be balanced by upward cross isopycnal mass flux in the ocean interior toward density ranges lighter than $\rho_1$. The interior fluxes are driven by small-scale mixing which depends on the vertical stratification there. If the surface cross-isopycnal flux is larger than the interior one, then the deep stratification must adjust, leading to accumulation of water mass in the deep ocean below the density $\rho_1$. This change in stratification would drive an adjustment in gradient-driven diapycnal mass fluxes, and  also change the surface outcrop location of $\rho_1$ and therefore the air sea fluxes and the surface transformation across this isopycnal. Thus the outcrop positions and deep stratification must co-evolve until a zero net flux across each isopycnal is achieved and an equilibrium is reached. The deep stratification set by this balance is exponential because, as was explained above, the solution of $w(z)\rho_z=(\kappa(z)\rho_{z})_z$ is very nearly exponential even if the upwelling or diffusivity are not constant in $z$. Within this interpretation, our findings are consistent with the recent studies pointing out the critical role of eddy processes and isopycnal slopes in the SO, except that the SO eddies in this picture communicate between the regions where cross-isopycnal water mass transformation occurs, but do not set the stratification on their own. Eddies, mixing and water mass transformation driven by air-sea fluxes all must be playing an important role in determining the exponential deep stratification.

The qualitative scenario suggested here, that the interior mixing balancing the surface water mass transformation occurs near ocean boundaries, is consistent with tracer release experiments showing weak interior small-scale mixing, as well as with the observed exponential profile of both the density and the buoyancy frequency. The picture considered here most straightforwardly applies as an explanation for the exponential density profile in the density range that outcrops in the SO only. However, the principle that the basic stratification is set by the balance of surface and deep cross isopycnal water mass transformation processes may apply also to the upper ocean thermocline stratification, including to inviscid thermocline theories that need to specify the basic stratification, typically on the eastern boundary \cite[][]{Luyten-Pedlosky-Stommel-1983:ventilated, Rhines-Young-1982:theory}, as was pointed out in \cite{Tziperman-1986:role}.

There are numerous idealizations used here, and many quantitative issues regarding the role of boundary mixing that should be further explored with closer examination of observations. The current study should therefore be viewed merely as an idealized exploration of processes that can lead to a deep exponential stratification, rather than a quantitative or definite explanation of the observed stratification.

\section{Acknowledgments}
MDM and ET were supported by NASA ROSES grant NNX14AH39G and NSF Physical Oceanography grant OCE-1535800. XY was supported by a summer undergraduate research fellowship from Peking University. Computational resources were provided by the NASA High-End Computing (HEC) Program through the NASA Advanced Supercomputing Center at NASA Ames and the NASA Center for Climate Simulation (NCCS) at Goddard Space Flight Center and on the Odyssey cluster supported by the FAS Division of Science, Research Computing Group at Harvard University. ET thanks the Weizmann Institute for its hospitality during parts of this work.

\newpage


\pagebreak

\begin{table}[t]
  \caption{Fit error measures between different temperature profiles shown in this paper and their exponential fits.
    W\&C corresponds to a calculated fit measure for the squared buoyancy frequency profile, calculated based on Fig.~13 of \cite{Wolfe2010}; N\&V similarly corresponds to Fig.~10 from \cite{Nikurashin2012}.}
  \label{table:fit-error}
\begin{center}
\begin{tabular}{ccccrrcrc}
\hline\hline
experiment & $T$ fit error (\%) & $T_z$ fit error (\%)\\
\hline
MinMix       & 1.852 & 2.547  \\
MunkMix      & 2.313 & 6.478  \\
MargMix      & 1.037 & 12.77  \\
W\&C CP-k1        & --   & 188.2 \\
W\&C CP-k8        & --   & 17.66 \\
N\&V Theory        & --   & 80.88 \\
N\&V Simulation      & --   & 105.91 \\
\hline
\end{tabular}
\end{center}
\end{table}

\clearpage
\newpage

\end{document}